\documentclass[useAMS,usenatbib]{mn2e}
\usepackage{graphicx,amssym}
\def\pp{{\prime\prime}}
\newcommand{\bc}{\begin{center}}
\newcommand{\ec}{\end{center}}

\title[Constraints on the star formation histories]
      {Constraints on the star formation
histories of galaxies from  $z \sim 1$ to $z \sim 0$}

\author[Chen et al.]{
\parbox[t]{\textwidth}{\raggedright
Yan-Mei Chen$^{1,2}$\thanks{Email: chenym@mail.ihep.ac.cn},
Vivienne Wild$^2$, 
Guinevere Kauffmann$^2$,
J\'{e}r\'{e}my Blaizot$^3$,
Marc Davis$^{4,5}$,
Kai Noeske$^6$,
Jian-Min Wang$^{1,7}$,
Christopher Willmer$^8$
}\\
\vspace*{6pt}\\
$^1$Key Laboratory for Particle Astrophysics, Institute of High Energy Physics, 
    Chinese Academy of Sciences, 19B Yuquan Road, \\
    Beijing 100049, China\\
$^2$Max--Planck--Institut f\"ur Astrophysik,
    Karl--Schwarzschild--Str. 1, D-85748 Garching, Germany\\
$^3$Universit\'e de Lyon, Lyon, F-69003; Universit\'e Lyon 1,
Observatoire de Lyon, 9 avenue Charles Andr\'e, Saint-Genis Laval,\\
    F-69230; CNRS, UMR 5574, Centre de Recherche Astrophysique de
    Lyon; Ecole Normale Sup\'erieure de Lyon, Lyon, F-69007, France\\    
$^4$Department of Astronomy, University of California, Berkeley, CA 94720, USA \\
$^5$Department of Physics, University of California, Berkeley, CA 94720, USA \\
$^6$W. M. Keck Postdoctoral Fellow at the Harvard-Smithsonian Center for
Astrophysics, 60 Garden St, Cambridge, MA 02138, USA\\
$^7$Theoretical Physics Center for Science Facilities, Chinese Academy of Sciences, China\\
$^8$Steward Observatory, University of Arizona, 933 N Cherry Ave, Tucson, AZ 85721, USA}
        
\begin{document}



\maketitle

\label{firstpage}

\begin{abstract}
We present a new method to estimate the average star formation
rate per unit stellar mass (SSFR) of a stacked population of
galaxies. We combine the spectra of $600-1000$ galaxies with similar
stellar masses and parameterise the star formation history of this
stacked population using a set of exponentially declining functions.
The strength of the Hydrogen Balmer absorption line series in the
rest-frame wavelength range $3750-4150$\AA\ is used to constrain the
SSFR by comparing with a library of models generated using the BC03
stellar population code. Our method, based on a principal component
analysis (PCA), can be applied in a consistent way to spectra drawn
from local galaxy surveys and from surveys at $z \sim 1$, and is only
weakly influenced by attenuation due to dust.  We apply our method to
galaxy samples drawn from SDSS and DEEP2 to study mass-dependent
growth of galaxies from $z \sim 1$ to $z \sim 0$.  We find that, (1)
high mass galaxies have lower SSFRs than low mass galaxies; (2) the
average SSFR has decreased from $z=1$ to $z=0$ by a factor of $\sim
3-4$, independent of galaxy mass.  Additionally, at $z \sim 1$ our
average SSFRs are a factor of $2-2.5$ lower than those derived from
multi-wavelength photometry using similar datasets.  We then compute
the average time (in units of the Hubble time, $t_{\rm H}(z)$) needed
by galaxies of a given mass to form their stars at their current
rate. At both $z=0$ and at $z=1$, this timescale decreases strongly
with stellar mass from values close to unity for galaxies with masses
$\sim 10^{10} M_{\odot}$, to more than ten for galaxies more massive
than $ 10^{11} M_{\odot}$. Our results are in good agreement with
models in which AGN feedback is more efficient at preventing gas from
cooling and forming stars in high mass galaxies.
\end{abstract}

\begin{keywords}
   galaxies: evolution -- galaxies: star formation
\end{keywords}

\section{Introduction}
\label{sec:intro}
Over the last decade, there have been a large number of  photometric and
spectroscopic surveys designed to study the formation and evolution of galaxies.
One major conclusion of these studies has been that the epoch when
massive galaxies formed most of their stellar mass is significantly earlier than   
that for low mass galaxies
\citep{Heavens04, Thomas05}. This phenomenon, popularly known as ``down-sizing'', 
at first sight seems  at odds with the predictions of   
the hierarchical CDM model, 
in which dark matter halos of all masses grow through merging and accretion
right up to the present day. The only way to reconcile the observations
with the theory, is to postulate that the growth of the most massive
galaxies is much slower than the growth of their surrounding halos.

The most natural way to achieve this is by invoking feedback
processes, which prevent gas from cooling, condensing and forming
stars in massive halos \citep{Bower06, Croton06, DeLucia06, Guo08}.  A
variety of different feedback mechanisms have been included in
numerical and semi-analytic models, for example feedback from quasars
and radio AGN \citep{Silk98, Hopkins06, wang06, Wang07}, supernova
heating \citep{Cole00, Benson03, Stringer08} and heating by infalling
substructure \citep{Dekel08}.  A physical understanding of feedback is
still lacking, however, and it is not understood which, if any, of the
proposed mechanisms are most important in regulating the growth of
galaxies.  It is likely that each mechanism will come into play at a
different mass scale and cosmological epoch.  By quantifying in detail
how galaxies of different masses grow as a function of time, we hope
to clarify how different galaxies form their stars and how this is
influenced by feedback processes.

There have been many attempts to measure the star formation rates of galaxy
populations from the present day out to redshifts greater than 5
\citep[see for example][]{Brinchmann04, Bauer05, Feulner05, Noeske07a}. 
For nearby galaxies, the H$\alpha$ line provides the most reliable estimate
of SFR, since it directly measures the number of ionizing photons
from massive stars. A reasonably reliable  correction for dust extinction
can be made provided one also measures the Balmer decrement
H$\alpha$/H$\beta$ accounting accurately for stellar absorption.
Beyond redshifts of $\sim 0.4$, the H$\alpha$ redshifts out of the
optical part of the spectrum and is no longer accessible. 
[O {\sc ii}] equivalent widths, ultraviolet (UV)-optical spectral 
energy distribution (SED)   fitting 
and infrared photometry are commonly used to estimate the star
formation rate. Each of these indicators is subject to different
disadvantages. The [O {\sc ii}] equivalent width is strongly affected by
dust and by  metallicity. 
The UV luminosities of galaxies are  also strongly affected by dust, while
the infrared only provides a direct measure of SFR if one has
measurements across the thermal peak of the spectrum. This has  not been  the case
for many of the Spitzer surveys aimed at quantifying
star formation in  high-redshift galaxies. All the indicators
discussed above  may also be
contaminated by AGN emission if there is an actively accreting
black hole in the galaxy.

In this paper, we develop an approach to  estimate the
amount of recent star formation experienced by a
{\em population} of galaxies. Our method is   based on the Balmer
absorption lines  located in the rest-frame wavelength 
range $3750-4150$\AA\ of the galaxy spectrum. 
The advantages of our method are the following: 

(1) The Balmer absorption lines are weakly influenced by 
dust attenuation and AGN contamination compared
with the indicators discussed above.   

(2) The wavelength range spanned by the Balmer absorption lines is accessible out
  to redshifts greater than 1, even in optical spectra. This means
  that the method can be applied in a consistent manner to both low 
  and high redshift ($z \sim 1$) samples.

(3) By stacking a large number of galaxies, we  estimate SFRs
  for a {\em complete sample} of galaxies in a given stellar mass range,
  including those galaxies that are forming stars weakly or not at all.
  These galaxies are often excluded when SFRs are measured
  for individual objects.  

We apply our method to a large sample of galaxy spectra from the
Sloan Digital Sky Survey (SDSS) and the DEEP2 redshift survey to study
the evolution of galaxies from $z=1$ to $z=0$. 
This redshift interval accounts for roughly half the age of the 
universe. Our study thus addresses the final stage of galaxy      
build-up in the Universe.

This paper is arranged as follows. In \S2, we introduce
the SDSS and DEEP2 samples used in our studies. The method to estimate
the amount of recent star formation in our galaxies is
developed in \S3. We apply the method to the SDSS and DEEP2 samples, and
present our results in \S4 and \S5, respectively. A discussion of the 
results is given in \S6. \S7 contains 
the summary of the paper. We use the cosmological
parameters $H_0=70~{\rm km~s^{-1}~Mpc^{-1}}$, $\Omega_{\rm M}=0.3$ and
$\Omega_{\Lambda}=0.7$ throughout this paper.

\section{Sample selection}

\subsection{The SDSS galaxy sample}
The low redshift galaxy sample comes from data release 4
\citep[DR4,][]{Adelman06} of the SDSS, 
which contains more than 550000 spectra in the primary
redshift range $0 \le z \le 0.3$. Stellar masses of the galaxies 
are obtained by fitting a suite of BC03 models 
\citep[described in][]{Salim07} to the SDSS model magnitudes. 
These masses are not identical to those of
\citet{Kauffmann03} or \citet{Gallazzi05}, who use spectral indices,
such as the D4000 and $\rm H\delta_A$ to constrain the mass-to-light
ratios of the galaxies. However, the differences between the mass
estimates are very small, the median offset of $\log M_*$ is only
0.01.  In this work we choose to use the photometric masses in order
to maintain consistency between our high and low redshift galaxy
samples. The masses assume the universal initial mass function (IMF)
as parametrized by \citet{Kroupa01}.

The aim of the first step of our work is to compare the SSFRs derived
from the Balmer series with the SSFRs of Brinchmann et al. (2004,
hereafter B04) derived from nebular emission lines.  We begin with a
subset of galaxies with $14.5<r<17.77$ and $0.005<z<0.22$ drawn from
the DR4 spectroscopic sample.  We divide galaxies with stellar masses
in the range of $10^{9} \sim 10^{12}M_\odot$ into 6 mass bins. The
mass interval is set to $\Delta \log M_{*}=0.5$.  The SDSS spectra are
obtained through a 3$^{\pp}$ circular fibre aperture, and therefore
sample primarily the inner regions of galaxies. To minimize the
effects of this so-called ``aperture bias'', we select galaxies with
$0.9<z/z_{\rm max}<1$, where $z_{\rm max}$ is the highest redshift at
which the galaxy in question would pass the sample selection
criteria. In Table 1, we list the mass range and median redshift of
galaxies in each mass bin.

\begin{table*}
\caption{The mass ranges and median redshifts of galaxies in the 6
mass bins of the SDSS sample and 4 mass bins of the DEEP2 sample.}
\begin{center}
{\footnotesize
\begin{tabular}{ccccccc}
\hline
$\log M_*/M_\odot=$ & $9.0-9.5$ & $9.5-10.0$ & $10.0-10.5$ & $10.5-11.0$ & $11.0-11.5$ & $11.5-12.0$ \\
\hline
$z$(SDSS) & $0.05$ & $0.07$ & $0.09$ & $0.14$ & $0.17$ & $0.20$ \\
$z$(DEEP2) & $-$ & $-$ & $0.85$ & $0.85$ & $0.85$ & $0.85$ \\

\hline
\end{tabular}
}
\end{center}
\label{tablefit}
\end{table*} 

\subsection{The high redshift galaxy sample}
In order to obtain accurate galaxy stellar masses at redshifts
approaching unity, photometry in both optical and near-IR passbands is required.
Our high redshift galaxy sample is selected from fields covered by both the
DEEP2 galaxy redshift survey and the Palomar Observatory Wide Infrared
Survey (POWIR). 

\subsubsection{The DEEP2 Survey}
The DEEP2 Galaxy Redshift Survey \citep[][ Faber et~al. in prep]{Davis03} utilizes the DEIMOS
spectrograph \citep{Faber03} on the KECK II telescope.  Targets for
the spectroscopic sample were selected from $BRI$ photometry taken
with the 12k x 8k mosaic camera on the Canada-France-Hawaii Telescope
(CFHT). The  images have a limiting magnitude 
of $R_{\rm AB} \sim 25.5$. As the $R$-band images have the
highest signal-to-noise ratio (S/N) of all the CFHT bands, they were
used to select the DEEP2 targets. The CFHT imaging covers four
widely-separated regions, with a total area of 3.5 $\rm deg^2$. In
fields 2-4, the spectroscopic sample is preselected using ($B-R$) and
($R-I$) colors to eliminate objects with $z<0.7$
\citep{Davis03}. Color and apparent magnitude cuts were also applied
to objects in the first field, the Extended Groth Strip (EGS), but
these were designed to downweight low redshift galaxies rather than
eliminate them entirely \citep{Willmer06}.  The third data release of
the DEEP2 survey\footnote{http://deep.berkeley.edu/DR3/} contains
spectra of about 50000 galaxies in the magnitude range $18.5 \le
R_{\rm AB} \le 24.1$. The spectra have a resolution of about $R \sim
5000$.

\subsubsection{The Palomar Observatory Infrared Survey}
$K_s$-band photometry was obtained in portions of all four fields
targeted by the DEEP2 galaxy redshift survey using the Wide Field Infrared
Camera (WIRC, Wilson et al. 2003) on the 5m Hale Telescope at the
Palomar Observatory.  \citet{Bundy06} mapped the central third
of fields 2-4 using contiguously spaced pointings tiled in a $3 \times
5$ pattern.  About 85\% of the pointings have a depth greater than
$K_{\rm AB}=22$ and the imaging covers a total of 0.9 $\rm deg^2$. In
field 1 (the Extended Groth Strip), the $K_s$-band imaging covers 0.7
$\rm deg^2$ to a typical depth of greater than $K_{\rm AB}=22.5$.

We use the stellar masses calculated by \citet{Bundy06}. They
estimate the $K_s$-band mass-to-light ratios for the galaxies in this
sample by comparing the optical to infrared spectral energy
distributions (SEDs) to a grid of 13400 model SEDs generated using the
BC03 code \citep{Bruzual03}.  The typical error in the stellar mass
estimates is about 0.2dex.  A
\citet{Chabrier03} IMF is used: the systematic difference between the
SDSS and DEEP2 stellar masses caused by the different choice of IMF is
about 0.05dex. We correct for this by adding 0.05dex to the
logarithm of the DEEP2 galaxy masses.

\subsubsection {Our high redshift galaxy sample}
The high redshift galaxy sample used in this work combines data from the
DEEP2 and POWIR survey. The criteria used to select the galaxies are the following:

\begin{enumerate}

\item The redshift range is $0.75 \le z \le 1$.  This ensures that all
the galaxy spectra  fully cover the  rest-frame
wavelength range required to measure the  spectral indices analyzed
in this paper  ($3750-4150$\AA) (Note that the  
DEEP2 spectra span the wavelength range $6500-9100$\AA).

\item The $R$-band magnitude range  is $18.5 \le R_{\rm AB} \le
  24.1$.  $R_{\rm AB} = 24.1$ is the limit of the DEEP2 spectroscopic
  sample.  By comparing with a fainter sample with photometric
  redshifts, \citet{Bundy06} concluded that the spectroscopic sample
  is essentially complete down to $M_* \ge 10^{10}M_\odot$.  As
  described in \S4.1 we will correct for any residual incompleteness
  by applying a $1/V_{\rm max}$ weight to each galaxy.

\item The $K_s$-band magnitude limit is $K_{\rm AB} \le 22$.  
The POWIR $K_s$-band survey covers different areas to different
depths. $K_{\rm AB} \le 22$ is chosen to maximize the sky area
covered, but also maintain a high level of completeness for galaxies
with stellar masses $\log M_{*}/M_\odot \ge 10.0$. This $K_s$-band
limit eliminates around 15\% sources with $10 \le
\log M_{*}/M_\odot \le 10.2$ from the $R_{\rm AB}\le 24.1$ sample. We
show in \S4.1 that this does not affect our results.

\end{enumerate}

In summary, our DEEP2 sample has a redshift range $z \sim 0.75-1.0$,
with $18.5 \le R_{\rm AB} \le 24.1$ and $K_{\rm AB} \le 22$.  It
covers a total area of 1.6 $\rm deg^2$ and the total source number is
about 3000. The galaxies have a mean and median redshift of 0.87 and
0.85 respectively.

\section{The spectral indices}

The 4000\AA\ wavelength region of the galaxy spectrum contains abundant
information for constraining the recent star formation
histories of galaxies.  Two traditionally-used
indices, D4000 and $\rm H\delta_A$, are located in this
region. Importantly,  this rest-frame wavelength region is usually
included in both low and intermediate redshift galaxy surveys.

PCA is a standard multivariate analysis technique, designed to
identify correlations in large datasets.
Using PCA, Wild et al. (2007, hereafter
W07) developed a set of new high signal-to-noise ratio spectral
indicators located in the rest-frame wavelength range
$3750-4150$\AA. At wavelengths around 4000\AA\, galaxy
spectra vary in both spectral shape and strength of the Hydrogen
Balmer absorption lines. These two are inversely correlated: younger
galaxies have stronger Balmer absorption lines and weaker 4000\AA\
break strengths. The third axis of variation is the Ca{\sc II}(H\&K)
lines, which are related to both the age and metallicity of the
galaxy.

Because the DEEP2 spectra are not flux calibrated, the continua
contain no useable information.  In this work, we do not use the
spectral indicators given by W07 but closely follow their method to
create a similar set of indicators designed to work on high
pass-filtered spectra.  Because the continuum is removed by the filter
process, we expect our eigenspectra to reveal the Balmer absorption lines as a
primary axis of variation and Ca{\sc II}(H\&K) lines as the secondary
axis.  We refer the reader to W07 for more details on the PCA
method. Here we present the dataset and the methods used to obtain the
indices and the resulting eigenspectra.

\subsection{Input model data set for PCA} 

Our input data set for the creation of the PCA eigenspectra is a set of
model spectra generated using the BC03 stellar population synthesis
code \citep{Bruzual03}. The model library is similar to that used in
\citet{Kauffmann03} and \citet{Salim05}, although with a more restricted 
parameter range. 6641 model galaxies are selected at random 
from the parent library 
according to the following 
criteria:

\begin{enumerate}
\item The time $t_{\rm form}$ when the galaxy begins to form its
stars is distributed uniformly between 0 and
5.7 Gyr after the Big Bang (the age of the universe is assumed
to be 13.7 Gyr).
\item The model galaxies have   exponentially
declining star formation histories  ${\rm SFR} \propto \exp(-t/\tau)$ with
$\tau$ distributed uniformally between $1 \le \tau \le 1.4$
Gyr. 
\item Top-hat bursts are superimposed on these continuous models. 
Two parameters describe the bursts: $f_{burst}$, the
fraction of the total stellar mass formed in bursts, is 
distributed logarithmically between 0.0 and 0.1;  
$t_{burst}$, the duration  of the burst, is
distributed uniformly between 0.03 and 0.3 Gyr. 
During the burst, stars form at constant rate. Bursts occur with 
equal probability at all times after $t_{\rm form}$ and the probability is 
set so that 50\% of the galaxies in the library have experienced a burst 
over the past 2 Gyr.
\item The metallicity is distributed linearly in the range 
$0.5 \le Z \le 2Z_\odot$; no metallicity evolution is included.
\end{enumerate}

These specific criteria (e.g. the distribution of $t_{form}$) are
chosen so that the model galaxies span a reasonably wide and even
range in mean stellar age. In this case, the Balmer absorption lines
and CaII(H+K) absorption lines will be isolated in the output
eigenspectra, and the Balmer absorption lines will dominate the first
eigenspectrum and CaII(H+K) absorption lines the second
eigenspectrum. We note that although changes to the input library will
alter the resulting eigenspectra, our calibration using
BC03 models in Section 5 ensures that we will recover the same SSFRs.

We convolve the model spectra  to have a velocity dispersion
equal to 150\,km/s; this is the median value of the velocity dispersion
of the galaxies in our samples.
Each model spectrum is then normalised by the mean flux in the rest-frame wavelength
range $4000-4080$\AA, where the spectrum is free from strong absorption
lines. 

We smooth the spectrum $F_\lambda$ with a Gaussian kernel
$W(\lambda)$. This yields the low-pass spectral component
\begin{equation}
F^{LP}_\lambda=\frac {\int d\lambda' W(\lambda-\lambda') Ivar(\lambda') F_{\lambda'}} {\int d\lambda' W(\lambda-\lambda') Ivar(\lambda')}
\end{equation}
where
\begin{equation}
W(\lambda-\lambda')=\frac{1}{\sqrt{2\pi}\sigma}\exp \left [-\frac{1}{2} \left (\frac{\lambda-\lambda'}{\sigma} \right )^2 \right ],
\end{equation}
and $\sigma=32$\AA. $Ivar(\lambda')$ is the inverse variance matrix of
the spectrum. For the model spectrum, it is equal to 1 at all
wavelength points. In the filtering process the strongest absorption
lines are masked by setting $Ivar(\lambda')=0$.  
We mask $\rm 20$\AA\ centred on 
$\lambda$4103, 3889, 3835; $\rm 16$\AA\ centred on 
$\lambda$3798; $\rm 12$\AA\ centred on $\lambda$3770, 3750. The sizes
of these masks were chosen to ensure that the absorption lines were
not diluted in the filter process. Small changes to the mask sizes make
no difference to the final results.

The high-pass component is then obtained as
\begin{equation}
F^{HP}_\lambda=F_\lambda-F^{LP}_\lambda.
\end{equation}
We choose to subtract the low-pass component, rather than divide by it as is commonly done,
because we wish to retain the luminosity weighting of the
real galaxies when they are stacked (see the following
Section). Figure 1 shows three examples of the filtering process. 

\begin{figure}
\bc
\hspace{-0.6cm}
\resizebox{8.5cm}{!}{\includegraphics{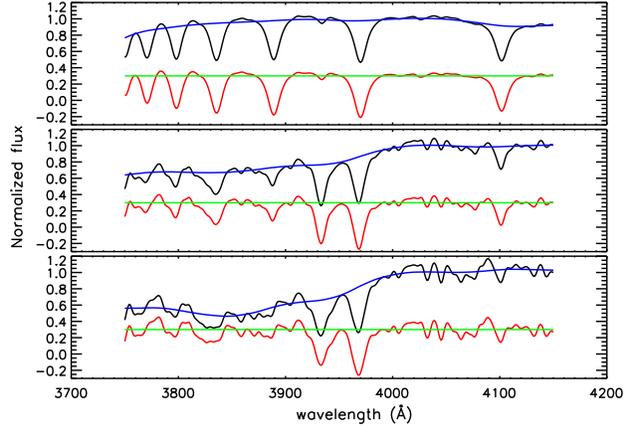}}\\%
\caption{Examples of the filter process applied to the model spectra. Black: the original 
model spectrum. Blue: the low-pass filter component. Red: the
resulting continuum filtered spectrum, shifted
upwards for easy comparison with the original spectrum. Green: the zero point. 
Models have been selected to cover the full range of galaxy ages:
galaxy age increases from top to bottom.}
\ec
\end{figure}

\begin{figure}
\bc
\hspace{-0.6cm}
\resizebox{8.5cm}{!}{\includegraphics{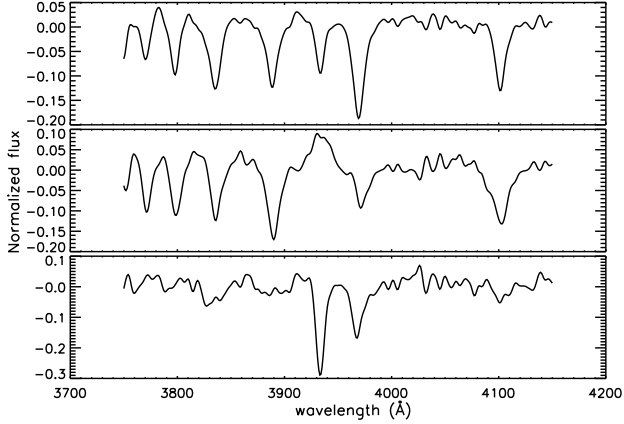}}\\%
\caption{Top: the mean spectrum of the continuum filtered input data set. Middle: 
the first eigenspectrum in which  the entire Balmer absorption line series is visible. 
Bottom: the second eigenspectrum, which shows the Ca{\sc II}(H\&K) absorption lines.}
\ec
\end{figure}

\subsection{Creating the `eigenspectra'}

The model library is dominated by old galaxies. To ensure the
dominance of the Balmer absorption lines in the first eigenspectrum,
we weight each model spectrum by the reciprocal value of its 4000 \AA\
break strength to reduce the impact of galaxies without strong Balmer
absorption lines. The mean spectrum of the input data set is then
calculated and subtracted, and the PCA is run to produce the
eigenspectra.

Figure 2 presents the mean spectrum and first two eigenspectra of our
input model galaxies. As required, the first eigenspectrum shows the
Balmer series. The Ca{\sc II}(H\&K) absorption lines are isolated in
the second eigenspectrum. In the following section the first 10
eigenspectra are used to fit the SDSS and DEEP2 spectra. The
coefficient of the first eigenspectrum ($PC1$, see eq.14), which
represents the strength of Balmer absorption lines, is used to derive
SSFRs.

\section{Data analysis} 
In this section, we describe how we apply our method to
real data. Our methodology is similar to that of 
\citet{Glazebrook03}, who stacked a large number of SDSS spectra to create
a ``cosmic average'' spectrum of all the stars in the Universe. The main difference is
that we carry out the cosmic averaging in bins of stellar mass.

The steps of the analysis are the following:
 
\begin{itemize}
\item The galaxy spectra are corrected for foreground Galactic 
extinction and shifted to the restframe.
\item The DEEP2 spectral resolution is about 3 times higher than that
  of the SDSS and the BC03 population synthesis  models.  In order to analyse
  SDSS and DEEP2 spectra in an equivalent manner, the DEEP2 spectra are
  convolved with a Gaussian kernel to match the resolution of the SDSS
  spectra.
\item The spectra are rebinned to match the model wavelength binning. 
\item The spectra are converted into units of luminosity (see below
  for details).
\item The continuum of the spectra is subtracted using the same filtering
method used  for the model spectra. An extra mask including 
$\rm 14$\AA\ centered on Ne {\sc iii}\,$\lambda$3783 line is added.
\end{itemize}

All the filtered spectra in the same mass bin are stacked using
the following weighting scheme: 
\begin{equation}
L_{\rm comp}(\lambda)=\frac{\sum_{i=1}^n {W(i)\times Ivar(i,\lambda) \times  L(i,\lambda)}}{\sum_{i=1}^n {W(i)\times Ivar(i,\lambda)}}
\end{equation}
where the sum is over $i=1,2,...,n$ galaxies with individual luminosities
$L(i,\lambda)$ in units of $\rm erg\, s^{-1}
{\rm \AA}^{-1}$. $Ivar(i,\lambda)$ is described below, and accounts for bad
pixels and night sky lines\footnote{For the DEEP2 spectra we make use
of the code {\sc coadd\_spectra.pro} available from the DEEP2 Team to identify
bad pixels and separate sky from continuum noise.}. $W(i)$ provides a
weight for each source to correct for missing galaxies due to survey
incompleteness.

For the flux calibrated SDSS spectra we convert to luminosity by
scaling by the luminosity distance of the galaxies in the standard
way.  We set $Ivar(i,\lambda)=0$ for bad pixels (identified in the
SDSS mask array), otherwise $Ivar(i,\lambda)=1$.  The redshift
intervals have been selected so that the samples are complete in mass
and we therefore set $W(i)=1$ for all galaxies. 

For the DEEP2 galaxies, the spectra are not flux calibrated and cannot
be converted directly into luminosity. In order to get $L(i,\lambda)$,
we first normalise each individual spectrum over all wavelengths to
its mean flux between $4000-4080$\AA. This normalised spectrum is then
scaled according to the optical photometry using the known
spectroscopic redshift and optical colour information. The
K-correction is derived from the best-fitting SED template output by
the {\sc KCORRECT} code \citep{Blanton07}.  The inverse variance
matrix $Ivar$ includes only variations due to sky intensity, not
object intensity$^2$. This ensures that bad pixels are downweighted
while equal weight is given to each spectrum.

Setting $Ivar(i,\lambda)=1$ and down-weighting the noisy pixels give similar results for SDSS. 
But for DEEP2, these two methods give very different results. In order to 
get the same S/N of the composite spectra, we need to stack many
more galaxies if we set $Ivar(i,\lambda)=1$ than if we down-weight the pixels
effected by sky lines in the DEEP2 spectra. This is because local
galaxy spectra (SDSS) have much much higher S/N than high redshift
galaxies (DEEP2) on average. The reason that individual pixel noise
arrays are not used in the $Ivar$ array, is that this would overall
downweight fainter galaxies, possibly introducing biases that are
difficult to quantify into the composite spectra.

\subsection{DEEP2 completeness correction}

An issue with every data set is the selection of weights to correct 
for missing galaxies. The weights we use in this study take into 
account three factors:

\begin{itemize}
\item The target selection strategies of DEEP2. 
\item The redshift success rate.  
\item The magnitude limits.
\end{itemize}

The weight $W(i)$ for DEEP2 galaxy $i$ is expressed as:
\begin{equation}
W(i)=\frac{\kappa_i}{V^i_{\rm max}}
\end{equation}
where $\kappa_i$ accounts for incompleteness resulting from the
DEEP2 color selection and redshift success rate. $V^i_{\rm max}$
accounts for the fact that faint galaxies are not detected throughout
the entire survey volume in a magnitude-limited survey.

To calculate $\kappa_i$ we closely follow the method described in
\citet{Willmer06}, but add an extra dimension to the reference data
cube, which is the $K_s$-band magnitude. For each galaxy in the
photometric catalogue, we select all the galaxies from the redshift
catalogue sharing the same bin in the $(B-R)/(R-I)/R_{\rm AB}/K_{s}$
space, count the number of objects with failed redshifts ($N_f$); the
number of galaxies with $z<z_{l}$ ($N_{z_{l}}$); the number of
galaxies with $z>z_{u}$ ($N_{z_u}$); and the number of galaxies with
successful redshifts within the ``legal" redshift range
$z_{l}$-$z_{u}$ ($N_z$). For field 1 (the EGS), $z_{l}$=0.2,
$z_{u}$=1.4; for fields 2-4, $z_{l}$=0.7, $z_{u}$=1.4.  The
probability that each galaxy in the photometric catalogue has a
redshift in the legal range is estimated.  For galaxies with
high-quality redshifts, the probability $P(z_{l} \le z \le z_{u})$=1
when $z_{l} \le z \le z_{u}$ and $P(z_{l} \le z \le z_{u})$=0 if
$z<z_{l}$ or $z>z_{u}$. The estimation of the probability for
unobserved sources is based on the so-called ``optimal" weighting
model (see \S2 of Willmer et al. 2006), which assumes that the failed
redshifts of red galaxies have the same distribution as the successful
ones, while blue galaxies with failed redshifts lie beyond the
redshift limit $z_u$. Namely, for the red galaxies:
\begin{equation}
P(z_{l} \le z \le z_{u})=\frac{N_z}{N_z+N_{z_{l}}+N_{z_{u}}} 
\end{equation}
for the blue galaxies
\begin{equation}
P(z_{l} \le z \le z_{u})=\frac{N_z}{N_z+N_{z_{l}}+N_{z_{u}}+N_f} 
\end{equation}
Finally, $\kappa_i$ is calculated by summing over all galaxies $j$
in the photometric catalogue with the same $(B-R)/(R-I)/R_{\rm AB}/K_{s}$
value as galaxy $i$, the probabilities that
the redshifts of galaxies are within the legal limits 
\begin{equation}
\kappa_i=\frac{\sum_{j}{P(z_{l} \le z \le z_{u}})}{N_z} 
\end{equation}

In the case of EGS, a final correction is applied to $\kappa_i$ to
account for the different sampling strategy, which
includes low-redshift ($z < 0.7$) galaxies but downweights them so
that they do not dominate the sample. The 
correction factor ($f_m$) depends on the location of the galaxy in
$(B-R)$ versus $(R-I)$ space and its apparent magnitude. $f_m$ has
been computed for around a quarter of the 
EGS galaxies from \citet{Willmer06} for the purpose of measuring
the galaxy luminosity function.  We estimated $f^i_m$ for the remaining
galaxies in the EGS data to be the average $f_m$ of all the galaxies
within the same bin in the $(B-R)/(R-I)/R_{\rm AB}$. The probability
that a galaxy will be placed on an EGS mask is given by
\begin{equation}
P({\rm mask})=0.33+0.43 P_{\rm gal} f_m 
\end{equation}
where $P_{\rm gal}$, given in the DEEP2 catalogue, is the probability
that an object is a galaxy based on its magnitude, color and size. For
EGS galaxies,
\begin{equation}
\kappa_i=\frac{\sum_{j}{P(z_{l} \le z \le z_{u}})}{N_z P(\rm mask)} 
\end{equation}
where $j$ includes all the galaxies in the photometric catalogue 
which are within $(B-R)/(R-I)/R_{\rm AB}/K_{s}$ space.

The calculation of $V^i_{\rm max}$  follows
\citet{Schmidt68}, and provides a simple way to account for the
$R$-band limit of the sample. We define
\begin{equation}
V^i_{\rm max}=d\Omega \int_{z_{min,i}}^{z_{max,i}} \frac{dV}{dz}dz
\end{equation} 
where $d\Omega$ is the solid angle  covered by the sample,  $\frac{dV}{dz}$ is the 
comoving volume element and $z_{min,i}$ and $z_{max,i}$ are the 
low and high redshift limits within which galaxy $i$ 
can be detected given the $R$-band limit of the survey. 
They are given by:
\begin{equation}
z_{{\rm max},i} = \min \{z'_{\rm max}, z(M_i,m_u)\}, 
\end{equation}
\begin{equation}
z_{{\rm min},i} = \max \{z'_{\rm min}, z(M_i,m_l)\},
\end{equation}
where $z'_{\rm min}$ and $z'_{\rm max}$ are the redshift  limits
of  the sample,  $z(M_i,m_u)$ is the redshift  above which the galaxy  would be fainter than
the $R$-band magnitude limit of $24.1$  and $z(M_i,m_l)$ is the redshift below
which the galaxy becomes brighter than  
$R_{\rm AB} = m_l=18.5$.
We use the best-fitting SED templates output by {\sc KCORRECT} to
calculate $z(M_i,m_u)$ and $z(M_i,m_l)$. Because we are
analyzing galaxies in a relatively narrow redshift slice, 
no  evolutionary correction is applied. 

In order to check that we recover the correct galaxy weights, we
calculate the stellar mass function for the redshift bin $0.75-1$ and
compare it to the result from \citet{Bundy06}.  Figure 3 shows that
the two stellar mass functions are almost identical.

\begin{figure}
\bc
\hspace{-0.6cm}
\resizebox{8.5cm}{!}{\includegraphics{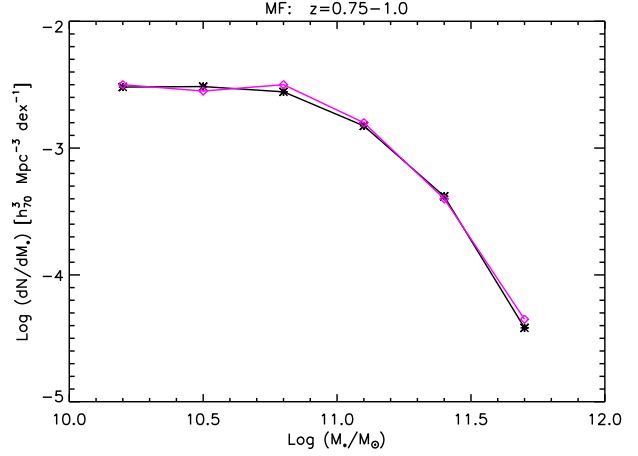}}\\%
\caption{The stellar mass function for galaxies in the redshift range $0.75-1$. The
black line is our result, the pink line is from Bundy et al. (2006).}
\ec
\end{figure}

In \S2.2.3, we noted that some sources are missing in our $\log
M_{*}/M_\odot=10-10.5$ DEEP2 mass bin due to the additional $K_{\rm AB}=22$ 
magnitude limit.  To test whether  this incompleteness affects our final
results, we repeat the entire analysis for a limiting magnitude 
$K_{\rm AB}=22.5$. This limit corresponds to a mass
completeness limit of about $10^{10}M_\odot$. We find that our measured
Balmer absorption line strengths in the lowest mass bin of DEEP2
remain the same to within the errors.   If we
were to increase the $K_s$-band magnitude limit to this fainter value
for the whole sample, the area covered by the sample would decrease and this
would affect our sample statistics, particularly for our highest mass bins.   

\subsection{Fitting composite spectra}

\begin{table*}
\caption{ Measured values of $H_{\rm balmer}$ and the SSFRs derived from them  
for each mass bin of SDSS and DEEP2. `S' represents SDSS, `D' represents DEEP2.}
\begin{center}
{\footnotesize
\begin{tabular}{lcccccc}
\hline
\hline
$\log M_*/M_\odot$=  & $9.0-9.5$  & $9.5-10.0$ & $10.0-10.5$ & $10.5-11.0$ & $11.0-11.5$ & $11.5-12.0$\\
\hline
$H_{\rm balmer}$(S) & $0.04 \pm 0.02$ & $-0.08\pm0.02$ &
$-0.20\pm0.01$ & $-0.34\pm0.01$ & $-0.46\pm0.01$ & $-0.53\pm0.01$ \\ 
$\log {\rm SSFR/yr^{-1}}$(S)& $-9.80\pm0.15$ & $-9.95_{-0.10}^{+0.15}$ & $-10.35_{-0.07}^{+0.15}$ &
$-10.75_{-0.15}^{+0.07}$ & $-11.15_{-0.17}^{+0.07}$ & $-11.60_{-0.55}^{+0.15}$ \\
$H_{\rm balmer}$(D) & $-$ & $-$ & $  0.03\pm0.04$ & $-0.11\pm0.03$ & $-0.21\pm0.03$ & $-0.46\pm0.04$ \\
$\log {\rm SSFR/yr^{-1}}$(D) & $-$ & $-$ & $-9.40\pm0.20$ & $-9.95_{-0.15}^{+0.20}$ & $-10.35\pm0.15$ & $-11.30_{-0.70}^{+0.20}$ \\

\hline
\end{tabular}
}
\end{center}
\end{table*}

\begin{figure*}
\bc
\hspace{-1.6cm}
\resizebox{17.cm}{!}{\includegraphics{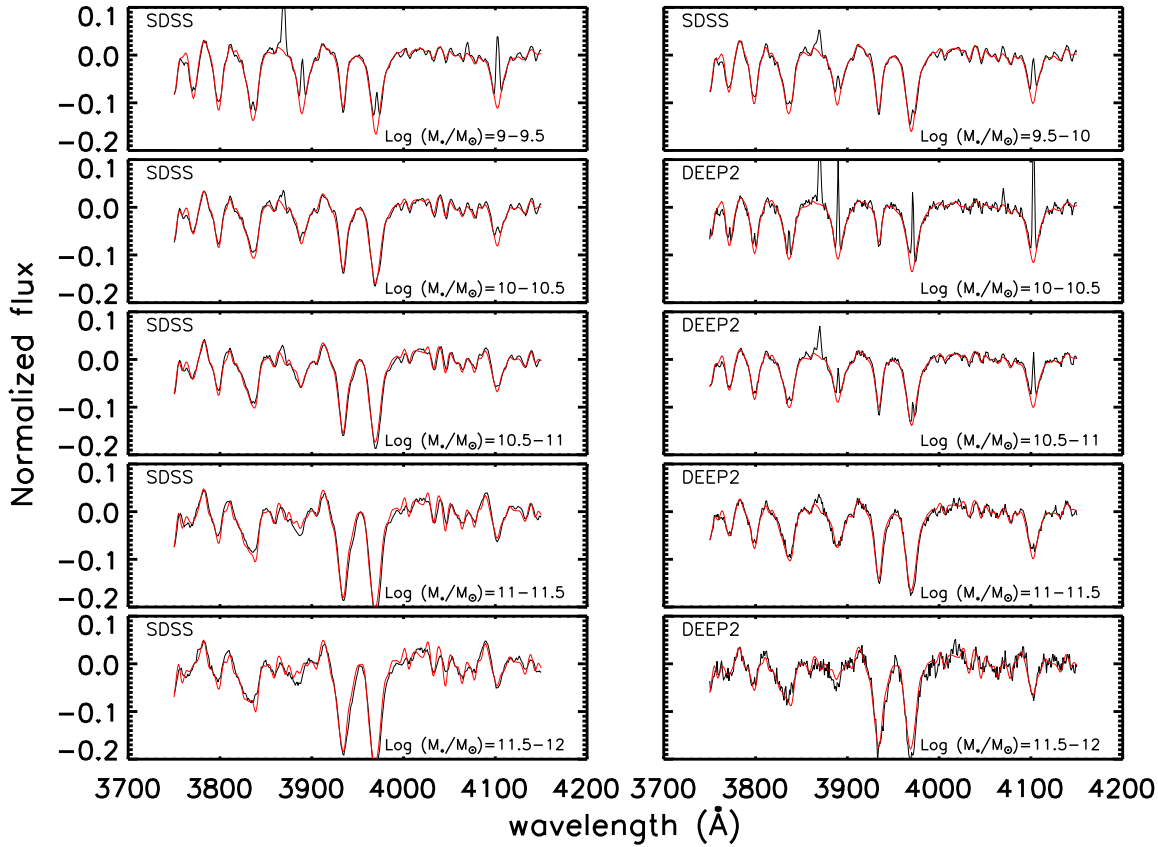}}\\%
\caption{The composite spectra (black) and PCA
fitted models (red, by eq.14) for all mass bins. The overall normalisation ($N$) has been 
divided out to make comparison easier.} 
\ec
\end{figure*}

The PCA projection algorithm employed here
accounts for pixel errors \citep{Connolly99} and allows the
normalisation of the spectrum as a free parameter (G. Lemson,
private communication, see W07). 
The fitting result is a linear combination of the eigenspectra, 
\begin{equation}
L_{\rm comp}(\lambda)=N \times \left(M+\sum_{i=1}^{10}PC_i\times E_i\right)
\end{equation}
where $N$ is the normalization, $M$ represents the mean spectrum
(Figure 2, top).  $PC_i$ is the amplitude of the $i-$th eigenspectrum
$E_i$, representing the amount of the $i-$th eigenspectrum present in
the residual spectrum. These amplitudes are the new spectral indices
which are calculated from the projection of each composite spectrum
onto the eigenspectra. We note that principal component amplitudes can
be negative, for example, negative $PC1$ means the strength of the
Balmer absorption lines in the composite spectrum is weaker than that
in the mean spectra. $E1$ represents the first eigenspectrum (Figure
2, middle), etc. In the current work, we are interested in the first
index, which measures the strength of the Balmer series. We refer to
it as `$H_{\rm balmer}$' from now on.  We would like to stress that
because we stack all spectra in each mass bin, the value of $H_{\rm
balmer}$ relates to the ``cosmic average'' spectrum, which includes
old galaxies with weak Balmer absorption lines.

Figure 4 shows the normalized composite 
spectra ($L_{\rm comp}(\lambda)/N$, black) and the the best-fit results (red). 
Both the strength of the Balmer absorption
and the nebular emission lines in these spectra reveal that for the
same mass bin, DEEP2 galaxies have much younger
populations than SDSS on average. The values of $H_{\rm balmer}$ are
listed in Table 2 for each mass bin of SDSS and DEEP2.  Although
statistical errors on the principal component amplitudes are
calculated during the projection of the eigenspectra, they are
insignificant with respect to the errors caused by the sample size. We
estimate the sample variance errors using a bootstrap technique, by sampling with
replacement the same number of galaxies as are in the stack, and
recalculating $H_{\rm balmer}$ 100 times. The errors are given by
the root-mean-square difference between the $H_{\rm balmer}$ values
calculated for each bootstrap sample and the mean
value.

\section{Estimating specific SFR}
When stacking many hundreds of galaxy spectra, it is reasonable to
assume that stochastic fluctuations in the star formation
histories of individual galaxies will average out, and that
the star formation history of the composite ``galaxy'' can be
approximated by a smooth function.  
In this section, we work with a grid 
of exponentially declining star formation models constructed from 
the BC03 population synthesis code. We create  $10^6$ different models
by varying the following parameters:

(1) The formation time of a galaxy ($t_{\rm form}$)
is defined as the time since the Big Bang when the 
galaxy first begins to form its
stars and is uniformly distributed between 0 and 10 Gyr.

(2) The star formation rate is modelled as  ${\rm SFR} \propto \exp(-t/\tau)$, 
with an e-folding timescale  $\tau$ uniformly distributed in
the range $0.01-10$ Gyr.

(3) We calculate three sets of models with metallicities
$Z=0.02=Z_\odot$, $Z=0.008$ and $Z=0.004$. According to the
mass-metallicity relation \citep{Gallazzi05, Panter08}, the average
global metallicity is near solar for the galaxies with stellar masses
greater than $10^{10}M_\odot$.  For galaxies with stellar masses
smaller than $10^{10}M_\odot$, most galaxies have sub-solar
metallicity, but the metallicity increases rapidly with stellar mass.
In the following, we will assume the observed mass-metallicity
relation and use solar metallicity models for mass bins with $M_* \ge
10^{10}M_\odot$, models with $Z=0.008$ for the mass bin with $9.5 \le
\log M_*/M_\odot \le 10$, and $Z=0.004$ models for the mass bin with
$9 \le \log M_*/M_\odot\le 9.5$.

(4) We consider models with and without dust. The effects of dust attenuation 
on the spectral properties are computed according 
to the simple two-component model of 
\citet{Charlot00}. The total effective $V$-band optical depth $\tau_{V}$ 
of the models is set to be a Gaussian distribution with peak at 1.6
and standard deviation of about 0.8. If $\tau_{V} <0$, we set it to
zero. This distribution is similar to the SDSS DR4 star forming
galaxies with stellar masses in the range of $10^{9}-10^{12} M_\odot$.
The parameter $\mu$, the fraction of the dust optical depth
contributed by the `ambient' interstellar medium, is set to be 0.3.
We will show later that our results are not greatly affected by dust.

We extract the spectrum for each of our model galaxies at an age
$age_{\rm model}=t_{\rm H}(z)-t_{\rm form}$, where $t_{\rm H}(z)$ is
the age of universe at redshift $z$, and $z$ corresponds to the median
redshifts of our SDSS and DEEP2 samples.  They are listed in Table 1.
After broadening the spectra to a velocity dispersion of 150km/s and
filtering out the continua, we measure the values of $H_{\rm balmer}$
for each of the model spectra.  The parameters of the models that best
fit the $H_{\rm balmer}$ values measured for the composite spectra,
are obtained by weighting each model in the library by the probability
function $\exp(-\chi^2/2)$, and then binning the probabilities as a
function of the given parameter value (see Appendix A of Kauffmann et
al.  2003 for a detailed description). The most probable value of a
model parameter for a particular composite spectrum can be taken as
the peak of this distribution; the most typical value is its median.
In the following sections, we use the median of the probability
distribution as our adopted best estimate.

\begin{figure}
\bc
\hspace{-0.6cm}
\resizebox{8.5cm}{!}{\includegraphics{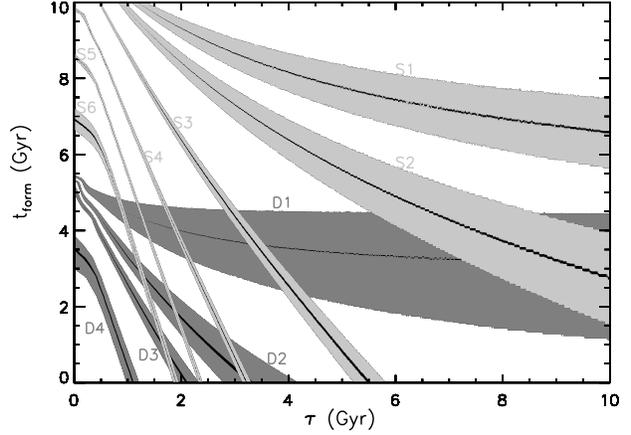}}\\%
\caption{Probability contours of $t_{\rm form}$ versus $\tau$ 
for galaxies in different stellar mass bins. They represent 
the 68\% confidence regions. Results are shown
for SDSS (light grey areas) and for
and DEEP2 (grey areas). 
The labels `S' and `D' in this plot represent SDSS and DEEP2, respectively. `S1' means the first
mass bin of SDSS, with $\log M_*/M_\odot=9.0-9.5$; `S2' with $\log M_*/M_\odot=9.5-10.0$; 
`S3' with $\log M_*/M_\odot=10.0-10.5$; `S4' with $\log M_*/M_\odot=10.5-11.0$; 
`S5' with $\log M_*/M_\odot=11.0-11.5$; `S6' with $\log M_*/M_\odot=11.5-12.0$; 
`D1' with $\log M_*/M_\odot=10.0-10.5$; `D2' with $\log M_*/M_\odot=10.5-11.0$; 
`D3' with $\log M_*/M_\odot=11.0-11.5$; `D4' with $\log M_*/M_\odot=11.5-12.0$.}
\ec
\end{figure}

Figure 5 shows our best-fit results 
in the space of  $\tau$ versus $t_{\rm form}$.
The shaded areas denote 68\% confidence regions with 
$|H_{\rm balmer}(m)- H_{\rm balmer}(d)| \le 1\sigma$, where 
$H_{\rm balmer}(m)$ is the value of $H_{\rm balmer}$ measured from the model 
galaxy spectrum. $H_{\rm balmer}(d)$ is measured from the composite spectrum, 
$\sigma$ is its error.
Light grey regions correspond to the 6 stellar mass bins for the SDSS sample,
while dark grey regions correspond to the 4 DEEP2 mass bins. The central black lines 
show the best-fit results.
As shown in Figure 5, there is a strong degeneracy between
$\tau$ and $t_{\rm form}$, so the 1-$\sigma$ contours appear as 
strips, rather than compact regions in the plot.
However, it is clear that the mass bins separate 
clearly in  $\tau$-$t_{\rm form}$ space, 
indicating different mass galaxies have had significantly different
{\em recent star formation histories.}

Rather than using our model grid to estimate the parameters $t_{\rm form}$
and $\tau$, which we have shown are degenerate, we now turn to a
parameter that directly characterizes the recent star formation
histories of the galaxies that make up our composite spectra. We
remind the reader that the Balmer absorption lines are most sensitive
to stars that have formed over the past $10^8-10^9$ years. In the
spectra of individual galaxies, the Balmer absorption lines have often been used
as a {\em post-starburst} indicator, because they remain strong for
nearly a Gyr after the burst has already ended. For our composite
spectra, however, bursts are not of relevance and our approach in this
paper will be to use $H_{\rm balmer}$ as a measure of the average
specific star formation rate (SSFR$=\langle \rm {SFR} \rangle/\langle M_*
\rangle$) of the galaxies in the stack.

We use the Balmer absorption lines as our SSFR indicator. These are mainly 
contributed by A stars (see Figure 1 of W07), which have a lifetime of 
several hundred million years. 
So long as the star formation rates of our composite galaxies are evolving
on timescales that are significantly longer than this, 
our estimate should agree with those that use shorter-timescale  indicators
such as emission lines or IR photometry. In the next subsection, we
will test this assertion using a catalogue of galaxy
spectra generated using a cosmological simulation. 

\begin{figure}
\bc
\hspace{-0.6cm}
\resizebox{8.5cm}{!}{\includegraphics{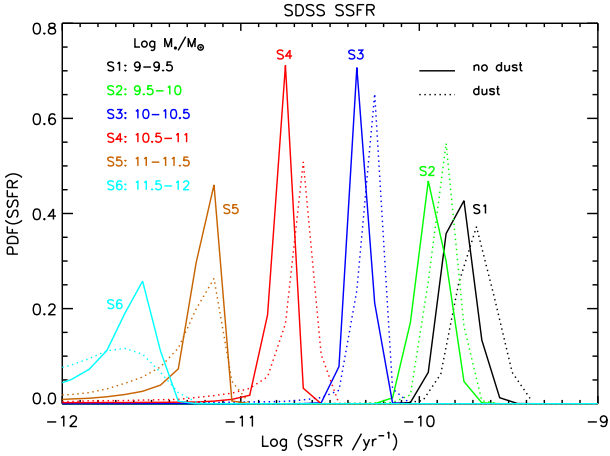}}\\%
\caption{The probability distributions of the SSFRs for SDSS galaxies. 
Lines with different colors represent different mass bins. 
The SSFRs decrease with increasing mass. Solid  lines show SSFRs 
estimated  using BC03 models without dust, while 
dashed-lines show the results using models with dust.}
\ec
\end{figure}

\begin{figure}
\bc
\hspace{-0.6cm}
\resizebox{8.5cm}{!}{\includegraphics{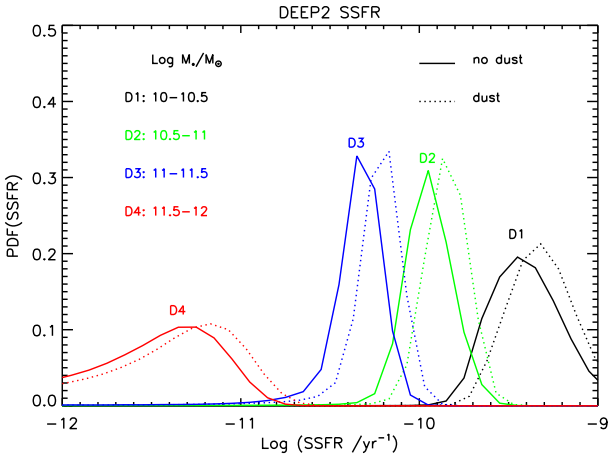}}\\%
\caption{Same as Figure 6, but for DEEP2.}
\ec
\end{figure}

In Figures 6 and 7, we show the probability distributions of the
parameter SSFR for each mass bin of the SDSS and DEEP2 galaxy samples.
The solid lines show SSFRs estimated using BC03 models without dust,
while dashed-lines show the results from models with dust.  As can be
seen, the results agree with each other to within 0.15dex, because
dust has a very weak effect on the Balmer absorption line series. From
now on, we will quote results based on models without dust.

In Figure 8, we plot the median values of the probability distribution
functions (PDF) of SSFR as a function of stellar mass for the galaxies
in the SDSS (green circles) and DEEP2 (red asterisks) samples.  The
error bars on each point indicate the 68 percentile range in the PDF
around the median. As can be seen the widths of the PDFs are quite
narrow, indicating that the formal error on our estimates of SSFR are
small. 

\begin{figure}
\bc
\hspace{-0.6cm}
\resizebox{8.5cm}{!}{\includegraphics{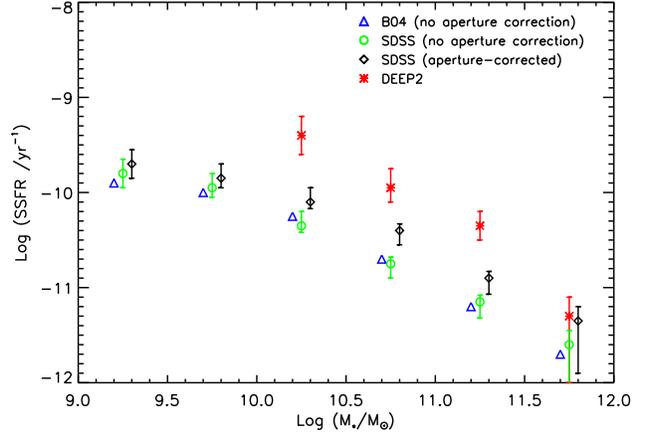}}\\%
\caption{SSFRs as a function of stellar mass. Green circles show the 
 SDSS results. Blue triangles show the fibre SSFRs estimated  by B04. 
Black diamonds show aperture-corrected SDSS SSFRs. The blue triangles and 
black diamonds have been slightly displaced in the x-axis (by minus and 
plus 0.05dex, respectively) to allow the error bars to be distinguished. 
Red asterisks: SSFRs for the four DEEP2 mass bins.}
\ec
\end{figure}

\subsection{Aperture effects}
When we compare the low redshift and high redshift measurements we
must account for aperture effects.

The SDSS spectra are obtained through 3$^{\pp}$ diameter fibres.  At
the median redshift of our sample, only the central 25\% of the galaxy
light has been included, so aperture effects will affect the SSFR
estimates, particularly for early-type spiral galaxies where the light
from the central few kpc will be dominated by the bulge component of
the galaxy and light from the disk will be underrepresented.  The
DEEP2 spectra are taken through 1$^{\pp}$ slits; the typical galaxy in
our high-$z$ sample has a diameter of $\le 2^{\pp}$.  Aperture effects
will thus be less of an issue here, unless SFR gradients are
significantly stronger at high $z$, which seems unlikely \citep{DP08}.
   
When we compare the SDSS and DEEP2 SSFR measurements in subsequent
sections, we must use aperture-corrected values for the SDSS
galaxies. We do this by applying the corrections estimated by
B04. They use the $g-r$ and $r-i$ colors measured outside the fibre to
estimate the amount of star formation in the outer parts of the
galaxies.  We note that
\citet{Salim07} showed that this correction is an overestimate
for certain classes of galaxies, such as AGN and galaxies without
detectable H$\alpha$ emission. Comparing the average
aperture-corrected SSFR of B04 in each mass bin with the results from
Salim et~al., we find that except for the most massive bin ($\log
M_*/M_\odot=11.5-12$), the discrepancy is negligible. We have thus
decided to apply the B04 correction to our 6 SDSS mass bins and the
results are shown in Figure 8 as black diamonds (they have been
displaced in the x-axis by plus 0.05dex to allow the error bars to be
distinguished). The aperture correction increases the value of SSFR by
$\sim 0.15$dex at $M_* \le 10^{10} M_\odot$ and $0.2
\sim 0.3$dex at $M_* > 10^{10} M_\odot$ We have not applied any
similar correction to the high-$z$ data. We will discuss this issue in
more detail in \S6.2.

Comparing our results for  SDSS and DEEP2 presented in Figure 8 leads to two main conclusions: 
 
\begin{enumerate}
\item At stellar masses above $10^{10} M_{\odot}$, in both 
the low redshift and the high redshift samples, the SSFRs 
decrease monotonically and relatively steeply  
with stellar mass (${\rm SSFR} \propto M_*^{-(0.85 \sim 0.9)}$) .

\item The average SSFR decreases by a factor 
of $\sim 3-4$ from $z \sim 1$ to $z \sim 0$, and this 
factor appears to be  independent of galaxy mass.
\end{enumerate}

\subsection{Tests of the method }
We have  claimed that our method of estimating the average SSFR
of a stacked sample of galaxies using $H_{\rm balmer}$ should agree
with other, more traditional methods that make use of indicators that
are sensitive to star formation over much shorter timescales.

In this section, we test this claim in two different ways.
Our first test is a {\em direct} one. B04 have estimated SSFRs
for the SDSS galaxies in our sample using emission line
fluxes (the greatest weight is carried by H$\alpha$).
For AGN and galaxies without detectable H$\alpha$ emission, 
the SSFR is estimated from the 4000 \AA\ break strength
measured within the fibre, using the relation between D4000 and SSFR calibrated 
using star forming galaxies. 

The blue triangles in Figure 8 show the average SSFRs from B04 for
exactly the same sample of galaxies used to derive the green circles.
We shift the blue triangles in the x-axis direction by minus 0.05dex
to make the comparison clear.  We find very good consistency between
the two methods, suggesting that $H_{\rm balmer}$ does provide a good
measure of the average SSFR of a population of galaxies and that our
methodology is robust. We note that the SSFRs derived by B04  from nebular
emission lines agree well with those derived from
multi-wavelength SED fitting including UV fluxes by Salim et
al. (2007). This provides yet another confirmation of the robustness of
the method.

We cannot test our estimates of SSFR for the DEEP2 galaxies in the
same way, because H$\alpha$ has shifted out of the relevant spectral
range.  Instead, we test our method using a DEEP2 mock catalogue of
galaxy spectra that has been generated from the Millennium Run
simulation \citep{Springel05}.  We first construct a $2\times 2$
sq. deg. mock catalogue using the MoMaF software \citep{Blaizot05}
and the semi-analytic prescription for galaxies from \citet{DeLucia07}
\footnote{The data for this model are publically available at
http://www.mpa-garching.mpg.de/millennium}.  
An important ingredient of this model, inherited from
the work of \citet{Croton06}, is the AGN feedback implementation,
which prevents cooling flows in massive haloes, and hence quickly
shuts down star formation in massive galaxies. In lower mass haloes,
where cooling continues unhampered by AGNs, the rate at which new cold gas 
reaches the galaxy and the efficiency of
galactic winds regulate the rate at which stars form. We
select mock galaxies to match the DEEP2/POWIR magnitude selection,
namely with $18.5 \le R_{\rm AB} \le 24.1$ and $K_{\rm AB} \le 22$.
We compute the stellar SED of each selected mock galaxy using its
complex star formation history and metallicity evolution predicted by
the semi-analytic model (SAM), combined with the BC03 stellar
population library. In the end, for each mock galaxy, we thus have an
``observed" SED plus a set of properties ($z$, $M_{*}$, SFR) taken
directly from the output of the SAM.  Although this catalogue may not
be a perfect representation of the real Universe, the co-moving star
formation rate density does increase at higher redshifts at a rate
similar to that observed (see Kitzbichler \& White 2007 for a
discussion).  The catalogue should thus provide a good way to test our
methodology using a sample of galaxies that is forming stars more
actively than those in the local Universe.

\begin{figure}
\bc
\hspace{-0.6cm}
\resizebox{8.5cm}{!}{\includegraphics{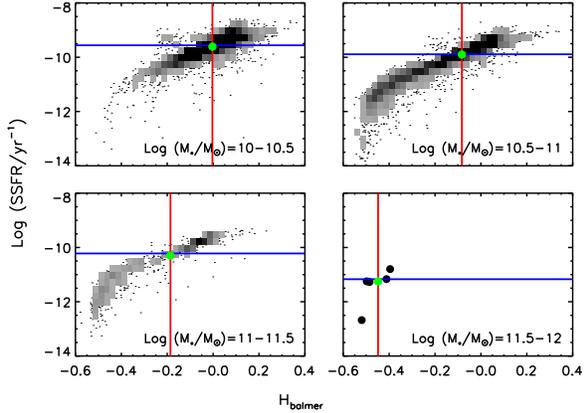}}\\%
\caption{SSFRs as a function of $H_{\rm balmer}$ for model galaxies in 
the DEEP2 mock catalogue with redshifts in the range 
$0.75 \le z \le 1$. Blue lines mark the mean SSFRs of galaxies in the mock catalogue 
evaluated directly from the stellar masses and SFRs predicted
by the semi-analytic model. Red lines mark the
values of $H_{\rm balmer}$ measured from the composite spectra.  The
green dots show the SSFRs derived from $H_{\rm balmer}$. The last panel has 
so few objects that they are shown as black points rather than greyscale.}
\ec
\end{figure}

As seen in Figure 1 of W07, the strength of Balmer absorption lines
decreases not only for old stellar populations, but also for the very
youngest stellar populations, which are dominated by O stars with weak
absorption lines.  If very young galaxies dominate the high redshift
samples, then one might worry that the method might underestimate the
specific star formation rates, simply because an O-star dominated
stellar population would be mistaken for a much older one.  We use the
DEEP2 mock catalogue to show that at redshifts $\sim 1$, the average
galaxy is not sufficiently young that this becomes a serious concern.
 
We have selected galaxies with $0.75 \le z \le 1$ from the DEEP2 mock
catalogue and divided them into mass bins in exactly the same way as
was done for the real data.  The analysis of the model galaxies
(including the stacking and filtering procedure) is also carried out
in the same way as for the real data.  In Figure 9, we plot SSFR as a
function of $H_{\rm balmer}$ for each individual model galaxy. Results
are shown for four different mass bins.  The blue lines show the
average SSFR of all the galaxies in the bin, defined as $\langle \rm
{SFR} \rangle/\langle M_* \rangle$. Red lines are the values of
$H_{\rm balmer}$ measured from the composite spectra.  The green dots
show the SSFRs derived from $H_{\rm balmer}$ of composite spectra. As
can be seen, the values obtained from the stacks are almost exactly
the same as those obtained from the average instantaneous SSFR
calculated from the model galaxies.

\section{Discussion}

In this section, we translate SSFR into a dimensionless star 
formation activity parameter (see Dav\'e 2008), defined as   
\begin{equation}
\alpha_{sf} \equiv \frac{1}{\rm SSFR}\frac{1}{t_{\rm H}(z)-1 \rm Gyr}. 
\end{equation}
Physically, this can roughly represent the fraction of the Hubble time (minus a
Gyr) that a galaxy needs to have formed its stars at its current rate
in order to produce its current stellar mass\footnote{We note that $\alpha_{sf}$ is 
not simply the fractional Hubble time the galaxy takes to form its
stars, because stellar mass is returned to the ISM, a fact which is
not accounted for in this simple model.}. A Gyr is subtracted in
order to take account of the fact that dark matter halos massive
enough to host galaxies with reasonably high star formation rates take
about 1 Gyr to assemble in a $\Lambda$CDM Universe \citep{Dave08}. 
A value of $\alpha_{sf}$ of 1 indicates that the galaxies could
feasibly have formed all their stellar mass by forming stars
continuously at the rate now observed. If $\alpha_{sf}$ is greater
than 1, their past average SFR must have been greater than their
current SFR for the stellar mass of the galaxy to have formed within a
Hubble time.

In Figure 10, we show our estimate of this characteristic timescale
$\alpha_{sf}$ as a function of stellar mass at both high (red
asterisks) and low (black diamonds) redshifts.  We remind the reader
that the low-$z$, highest mass bin is uncertain due to aperture
correction complications in the SDSS survey. From this plot, we find
that $\alpha_{sf}$ is a strongly increasing function of $M_*$; the
timescale increases from values close to a Hubble time for galaxies
with $10^{10} M_{\odot}$ to more than an order of magnitude larger
than the Hubble time for galaxies more massive than
$10^{11}M_{\odot}$.  We find that, at fixed stellar mass, $\log
\alpha_{sf}$ evolves only slightly with redshift, decreasing by around
0.2dex between $z\sim0$ and $z\sim1$.

\begin{figure}
\bc
\hspace{-0.6cm}
\resizebox{8.5cm}{!}{\includegraphics{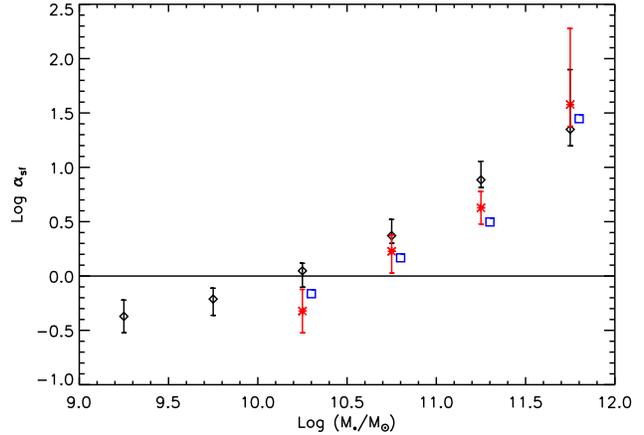}}\\%
\caption{Star formation activity parameter $\alpha_{sf}$ as a function of stellar mass. 
As in previous plots, black diamonds and red asterisks represent SDSS and DEEP2 results, 
respectively. Blue squares are the results derived from the DEEP2 mock catalogue, they have 
been displaced in the x-axis by plus 0.05dex to make the comparison clear.}
\ec
\end{figure}

\subsection{Comparison with galaxy formation models}

The blue squares in Figure 10 (displaced in the x-axis by plus 0.05dex
to make the comparison clear) show the characteristic timescale of
star formation in the mock DEEP2 universe created from the Millennium
run SAM as described in Section 5.2. A key result of this paper is
that the model is entirely consistent with observational results at
$z\sim1$, both in amplitude and slope.

More generally, in theoretical models of galaxy formation, at a fixed
stellar mass, $\alpha_{sf}$ is predicted to remain constant out to
redshifts greater than $2$ \citep[Figure 2 of][]{Dave08}. This latter result is
robust to methodology, both semi-analytic and
smooth-particle-hydrodynamic (SPH) simulations agree. Specifically,
between a redshift of 1 and 0 the models predict changes in
$\log\alpha_{sf}$ of less than 0.1dex. Our observational results
indicate a slightly larger increase of 0.2dex. Further
studies using larger surveys will be required to confirm this
small amount of evolution.

The amplitude of $\alpha_{sf}$ and its variation with stellar mass
depend on simulation methodology.
The near unity $\alpha_{sf}$ at both $z\sim0$ and $z\sim1$ for galaxies with
$M_* \la 10^{10.5}M_{\odot}$ suggests that galaxy mass growth in this
mass and redshift range may be dominated by smooth and steady cold mode
accretion, as implemented in all current models of galaxy formation. 
The strong increase of $\alpha_{sf}$ with $M_*$ that we observe at
both redshift 1 and in the local Universe (Figure 10), is an
expression of the phenomenon of `downsizing': massive galaxies have
apparently completed most of their star formation at higher redshifts
than low mass systems. In many current
models of galaxy formation, the explanation of this behaviour is ``AGN
feedback''.  More massive galaxies are more likely to host massive
black holes which have the capability of producing more energy.  Of
equal importance, these massive galaxies are hosted by larger halos,
where the AGN energy can be well coupled to the material that would
otherwise cool and fuel star formation in the galaxies.  As a result,
AGN feedback through heating of the interstellar and intergalactic gas
is more efficient in massive galaxies. 

In summary, the observed amplitude and evolution of  $\alpha_{sf}$
as presented in this paper, provide firm constraints for
all galaxy formation models.

\subsection{Comparison with previous results}

As we have shown, our results appear to be in good general agreement
with cosmological galaxy formation models. We now turn to a comparison
with previous results from the literature. As we shall show, there is
a significant discrepancy of $0.3-0.4$dex between our $H_{\rm Balmer}$
derived SSFRs, and those derived primarily from multiwavelength broad
band photometry.   

\begin{figure}
\bc
\hspace{-0.6cm}
\resizebox{8.5cm}{!}{\includegraphics{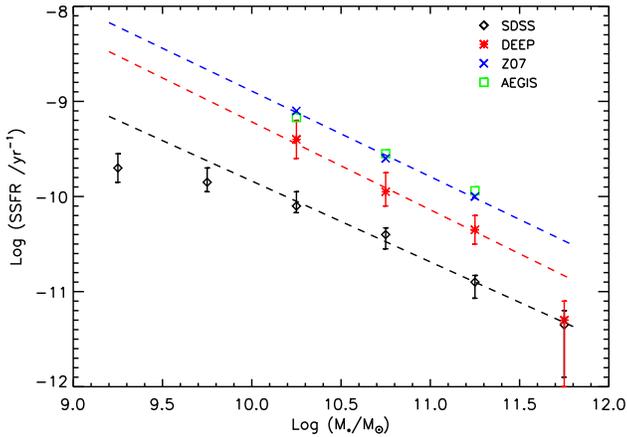}}\\%
\caption{SSFR as a function of stellar mass. Black diamonds: 
aperture-corrected SDSS data. Red asterisks: SSFRs of DEEP2 galaxies. 
Blue crosses and green squares: SSFRs from Z07 \& AEGIS respectively.}
\ec
\end{figure}

The results from two other studies of the SSFRs of galaxies at $z \sim
1$ are compared with our results in Figure 11.  The blue crosses
represent SSFRs derived from the COMBO-17 sample with Spitzer 24$\mu$m
and GALEX data (Zheng et al. 2007, hereafter Z07).  These authors use
the measured UV+IR luminosities to derive SFRs \citep{Bell05}. They
take account of the contribution from galaxies that are not
individually detected at 24$\mu$m by stacking their images.

The green squares show results calculated by us for a sample of
galaxies selected from the All-Wavelength Extended Groth Strip
International Survey (AEGIS). The galaxies are a subsample of those
used to create the composite spectra in this paper.  This sample is
not exactly the same as presented in Noeske et~al. (2007, hereafter
N07) and \citet{Noeske07b}, as only star-forming galaxies were
included in their analysis. Instead, we have averaged the SSFRs over
all the galaxies in each stellar mass bin (i.e. both star-forming and
quiescent ones), using the same weighting factors that we used in our
spectral stacking analysis. Aside from this difference in sample, the
method used to derive SFRs for the AEGIS galaxies is the same as used
by N07.

The method used by N07 to derive SFRs is slightly different to that
employed by Z07, in that information from emission lines in the spectra
of the galaxies is utilized if it is available. 
For galaxies with $f_{24\mu m}>60\mu$Jy and strong emission lines, 
the total SFR is derived from a combination of the IR measurements and 
from DEEP2 emission lines (H$\alpha$, H$\beta$, or [O {\sc ii}]$\lambda$3727, 
depending on $z$) with no extinction correction. SFRs are derived 
from extinction-corrected emission lines only  for blue galaxies 
with strong emission lines and no detectable 24 micron emission. 
Red galaxies with weak emission lines, but no 24 micron detections,
are considered star-forming and SFRs are derived from emission lines, 
assuming the same extinction corrections as for normal  
star-forming galaxies. 


The dashed lines in Figure 11 are a linear fit to the data points for
mass bins with $\log M_{*}/M_\odot \ge 10$.  As can be seen from this
plot, the slope of the relation between SSFR and mass that we derive
is consistent with the results of AEGIS and Z07.  However our method
yields a normalisation that is $0.3-0.4$dex lower than AEGIS and Z07.

This discrepancy is puzzling, but it does have a number of possible 
explanations:

1) {\em Systematic differences in the calibration of different SFR
indicators.} Our method is based on stellar absorption line
indicators, while the N07 and Z07 results are based on a combination
of UV, IR and emission lines. We have demonstrated that our results do
agree with the SSFRs derived by B04 from emission lines at low
redshift, which provides confidence that there is no significant
discrepancy between our method and the standard calibration of
extinction-corrected H$\alpha$ to derive SFR.  We are unable to carry
out the same test at $z \sim 1$, because H$\alpha$ is redshifted out
of the spectral range covered by most galaxy surveys. Systematic
differences in estimated stellar mass do occur when different
population synthesis models are used due to differing mass-to-light
ratios in the models. However, the same BC03 models have been used in
the comparison to the AEGIS results.

2) {\em Obscured AGN.} There has been no attempt to remove obscured
AGN from the two $z=1$ galaxy samples with which we compare. Both the
N07 and Z07 analyses make use of the 24 micron Spitzer passband as a
star formation indicator.  At $z=1$, this corresponds to a rest-frame
wavelength of 12 micron, which is very close in wavelength to where
emission from a dusty torus would become very significant \citep[see
e.g.][]{Daddi07}. In addition, AGN emission could well be
contaminating some of the optical emission lines used to estimate
SFR. By contrast, the $H_{\rm balmer}$ index originates from stellar
atmospheres and is not expected to be contaminated from emission from
an obscured AGN. 

On the other hand, the vast majority of DEEP2
24 micron sources and  line--emitting galaxies have
line ratios indicating star formation and not AGN
\citep{Weiner07}. 
If the AGN were highly obscured, they could contribute at 24 micron
and not show up in optical lines, but to make up a difference of
0.3dex one would have to assign half of the 24$\mu$m emission at z=1
to obscured AGN, which would appear quite extreme (B.~Weiner, private
communication).

3) {\em Evolution of IMF with redshift.} The SFR indicators used by
N07 and Z07 trace O and B stars, whereas our SFR indicator is the
Balmer series and is mainly influenced by A stars.  If the IMF changes
with redshift such that more massive stars form at higher $z$
\citep[e.g.][]{Dave08, Dokkum08}, then a discrepancy between the two
methods may not be apparent in the analysis of the low redshift
samples, but may become more pronounced at higher $z$.

4) {\em Aperture effects.}  As we have discussed, the SSFRs we
estimate for the $z=1$ galaxies may be biased somewhat low because the
long-slit spectra preferentially sample the inner bulge of the
galaxy. However, 95\% of the DEEP2 galaxies (at all redshifts) with 
a line measurement have $r_{eff} < 0.95^{\pp}$ as measured in the CFHT 
imaging \citep{Weiner07}.  Thus the 1$^{\pp}$ slit covers a large fraction of the galaxy. 
Additionally the seeing mixes the light into the slit to a much greater degree 
in DEEP2 than it does for the SDSS fibers. 
Thus the star formation gradients of $z=1$ galaxies would
have to be extremely strong for this to make a factor $2-3$ difference
to the SSFR estimated for the galaxy population as a whole. As can be
seen from Figure 8, the correction for aperture effects in the low $z$
sample is only a factor of 2 on average, even in the case where the
fibre only samples 25\% of the total light.

In this paper, we are not able make a definitive conclusion with regard to the 
possibilities listed above. It is clear that there are many inherent
uncertainties in estimating SFRs in galaxies and that more work is
needed before the factors of $2-3$ offsets that we see between the
different methods and the models can be understood in detail. 

\section{Summary}
In this paper, we developed a new method to measure the average SSFR
of a population of galaxies using the Balmer absorption line series
located in the rest-frame wavelength range $3750-4150$\AA. Our method is
free of complications due to dust extinction and AGN contamination,
and provides a consistent way to measure SSFRs at both high and low
redshifts. The robustness of the method has been tested using SDSS
data and a DEEP2 mock catalogue drawn from cosmological simulations.

We apply this method to the DEEP2 galaxy sample. We high-pass filter
the spectra because the DEEP2 spectra are not flux calibrated and then
stack together the spectra of galaxies with similar stellar masses.
SSFRs are estimated from the Balmer absorption line series by
comparing to a library of model spectra generated from the BC03
stellar population code.

Our results show that:
\begin{itemize}
\item  The average SSFR decreases monotonically with
stellar mass at both $z\sim0$ and $z\sim1$. At both redshifts the
decrease is almost an order of magnitude between $10^{10}M_{\odot}$
and $10^{11.5}M_{\odot}$.
\item For galaxies of fixed stellar mass, the average SSFR has
decreased by a factor of $3-4$ from $z\sim1$ to $z\sim0$. 
\item The amplitude of the decrease in SSFR with $z$ is independent of stellar
mass in the stellar mass range observable in the high redshift sample,
i.e. $10^{10}-10^{11.5}M_{\odot}$.
\item The average SSFR of galaxies at $z\sim1$ is consistent with the
predictions of the semi-analytic model of \citet{DeLucia07}
in all stellar mass bins observed.
\end{itemize}

We define a star formation activity parameter $\alpha_{sf}$ which is
the average time (in units of the Hubble time at redshift $z$) needed
by a galaxy of a given mass to form its stars at its current SFR. We
find that galaxies of mass $\la10^{10.5}M_{\odot}$ are consistent with
$\alpha_{sf}\sim1$. This is in good agreement with models in which
star formation is regulated by the infall of cold gas. At higher
masses $\alpha_{sf}$ increases sharply, which, in the currently
favoured galaxy formation scenario, can be understood by AGN feedback
reducing the rate at which gas cools onto more massive galaxies.

\section*{acknowledgements}

We are very grateful to the referee for useful comments and suggestions 
that have strengthened this work. We also thank 
Kevin Bundy and Jarle
Brinchmann for generously providing access to their data; Simon White
and Qi Guo for helpful discussions. YMC is supported by an EARA
fellowship. VW is supported by the MAGPOP Marie Curie EU Research and
Training Network. JMW is supported by NSFC 
and CAS via NSFC-10325313, 10733010, 10521001, KJCX2-YW-T03, and 973 project 
2009CB824800.

The low redshift data of this work comes from the Sloan Digital Sky Survey (SDSS).
Funding for the Sloan Digital Sky Survey (SDSS) has been provided by
the Alfred P. Sloan Foundation, the Participating Institutions, the
National Aeronautics and Space Administration, the National Science
Foundation, the U.S. Department of Energy, the Japanese
Monbukagakusho, and the Max Planck Society. The SDSS Web site is
http://www.sdss.org/.  The SDSS is managed by the Astrophysical
Research Consortium (ARC) for the Participating Institutions. The
Participating Institutions are The University of Chicago, Fermilab,
the Institute for Advanced Study, the Japan Participation Group, The
Johns Hopkins University, Los Alamos National Laboratory, the
Max-Planck-Institute for Astronomy (MPIA), the Max-Planck-Institute
for Astrophysics (MPA), New Mexico State University, University of
Pittsburgh, Princeton University, the United States Naval Observatory,
and the University of Washington.

This work is also based on observations with the W.M. Keck Telescope, the Hubble Space Telescope, 
the Canada France Hawaii Telescope, and
the Palomar Observatory, and was supported by NASA and NSF grants.

\bibliographystyle{mn2e}

\bibliography{evolution}

\end {document}